\documentclass[prb,reprint,superscriptaddress,citeautoscript]{revtex4-1}

\usepackage[applemac]{inputenc}
\usepackage[T1]{fontenc}
\usepackage[american]{babel}
\usepackage[scaled]{helvet}
\usepackage{graphicx,amssymb,bm,mathtools,textcomp,courier,multirow,hyperref,color,pdfpages,subfigure}
\usepackage[normalem]{ulem}	%fixes problem with book references

\newcommand{\Dirac}[3]{\left\langle #1 \left| #2\right| #3\right\rangle}

\newcommand{\ket}[1]{\left|\left. #1 \right\rangle\right.}

\newcommand{\figureshortname}{Fig.}
\newcommand{\equationshortname}{Eq.}
\newcommand{\tableshortname}{Tab.}

\newcommand{\eref}[1]{\equationshortname~\eqref{#1}}

\newcommand{\aref}[1]{Appx.~\ref{#1}}
\newcommand{\fref}[1]{\figureshortname~\ref{#1}}
\newcommand{\tref}[1]{\tableshortname~\ref{#1}}
\newcommand{\rcite}[1]{Ref.~[\onlinecite{#1}]}

\graphicspath{{Pictures/}}

\begin{document}
\def\sectionautorefname{Sec.}

\title{Noise-Protected Gate for Six-Electron Double-Dot Qubits}

\author{Sebastian Mehl}
\email{s.mehl@fz-juelich.de}
\affiliation{Peter Grünberg Institute: Theoretical Nanoelectronics, Forschungszentrum Jülich, D-52425 Jülich, Germany}
\affiliation{Institute for Quantum Information, RWTH Aachen University, D-52056 Aachen, Germany}

\author{David P. DiVincenzo}
\affiliation{Peter Grünberg Institute: Theoretical Nanoelectronics, Forschungszentrum Jülich, D-52425 Jülich, Germany}
\affiliation{Institute for Quantum Information, RWTH Aachen University, D-52056 Aachen, Germany}
\affiliation{Jülich-Aachen Research Alliance (JARA), Fundamentals of Future Information Technologies, D-52425 Jülich, Germany}

\date{\today}

%------------------------------------------------------------------------------------
%------------------------------------------------------------------------------------
%------------------------------------------------------------------------------------
\begin{abstract}
Singlet-triplet spin qubits in six-electron double quantum dots, in moderate magnetic fields, can show superior immunity to charge noise. This immunity results from the symmetry of orbitals in the second energy shell of circular quantum dots: singlet and triplet states in this shell have identical charge distributions. Our phase-gate simulations, which include $1/f$ charge noise from fluctuating traps, show that this symmetry is most effectively exploited if the gate operation switches rapidly between sweet spots deep in the (3,3) and (4,2) charge stability regions; fidelities very close to one are predicted if subnanosecond switching can be performed.
\end{abstract}

\maketitle

%------------------------------------------------------------------------------------
%------------------------------------------------------------------------------------
%------------------------------------------------------------------------------------
\section{Introduction}
The spin degree of freedom of the few-electron quantum dot (QD) is an excellent building block for a qubit. While a single electron spin may serve directly as a qubit \cite{loss1998}, the difficulty of single-qubit operations makes it desirable to encode a qubit in a multielectron state.  Considerable success has been achieved with a two-electron encoding \cite{levy2002}, in which the singlet and spinless triplet levels of the double quantum dot (DQD) define a logical qubit \cite{hanson2007-2}.  Electric pulses, applied on the microsecond scale, permit all necessary one- \cite{petta2005,johnson2005} and two-qubit \cite{weperen2011,shulman2012} operations when supplemented by magnetic field gradients \cite{foletti2009,gullans2010,brunner2011}. 

This paper addresses the crucial {\it exchange gate}, which has provided a route to impressive progress in the singlet-triplet qubit \cite{petta2005,johnson2005}.  In this gate a DQD is moved away from the ``neutral'' electron distribution, i.e.  having one electron on each QD [referred to as $\left(1,1\right)$] to one having a slight bias towards double occupancy of one QD [e.g. the left QD: $\left(2,0\right)$]. Only the singlet configuration permits electron transfer from $\left(1,1\right)$ to $\left(2,0\right)$, while transfer from the triplet state is blocked (Pauli spin blockade). Exchange gates allow fast qubit manipulations since they couple strongly to the charge instead of the spin degree of freedom. But new noise mechanism consequently emerge: low-frequency switching of charge traps become a major problem \cite{coish2005,hu2006,taylor2007}.

Here we show that, paradoxically, the exchange gate will be much less susceptible to change noise if the DQD is pulsed {\it fully} from the $\left(1,1\right)$ to the $\left(2,0\right)$ regime. Pulsing far into the $\left(2,0\right)$ region also lifts the spin blockade for the triplet state, as an excited orbital state becomes energetically accessible \cite{nielsen2010,dial2013}. The singlet-triplet splitting is highly protected from charge noise deep in the $\left(2,0\right)$ region.  We show that the fidelity of exchange gates will be excellent under two conditions: (1) the pulse rise and fall times should be subnanosecond, and (2) the electrons should be in the {\it second shell}, so that singlet and triplet states have the same charge distribution.  This means that the best exchange gate is predicted to occur for the six-electron DQD with four nonparticipating ``core'' electrons, so that the desired transition is actually between $\left(3,3\right)$ and $\left(4,2\right)$.

Presently only one other approach has been proposed to mitigate charge noise. There is the suggestion to encode singlet-triplet qubits into many-electron QDs ($N>3$), so that background electrons may screen charge fluctuations \cite{barnes2011,nielsen2013, higginbotham2014}. This approach involves strong renormalizations of the QD's one-particle wave functions when interacting with charge traps; our approach is quite distinct, involving only weak state renormalizations.

%------------------------------------------------------------------------------------
%------------------------------------------------------------------------------------
%------------------------------------------------------------------------------------
\section{Model}
Our description of DQDs starts with the single-particle eigenstates of a circular QD with confining potential $V\left(x,y\right)=\frac{m\omega_0}{2}r^2$ \cite{burkard1999,ferry2009} and out-of-plane magnetic field $B$. The eigenstates are the Fock-Darwin (FD) states 
$\psi_{n,l} \left(\rho,\phi\right)$ 
$=$ 
$\left(m\Omega/\hbar\right)^{\frac{\left|n\right|+1}{2}}$
$\sqrt{\frac{n!}{\pi\left(n+\left|l\right|\right)!}}$
$L_n^{\left|l\right|}\left[\left(m\Omega/\hbar\right)\rho^2\right]$
$e^{-il \phi} \rho^{\left|l\right|}$
$e^{- \frac{m \Omega}{2\hbar}\rho^2}$, with $n\in\mathbb{N}_{0}$ and $l\in\mathbb{Z}$. We use polar coordinates $\left(\rho,\phi\right)$, while $L_{i}^{j}\left(x\right)$ are the generalized Laguerre polynomials. States of the same energy shell all have the same value of $2n+\left|l\right|$, with energies $E_{n,l}=\left(2n+\left|l\right|+1\right)\hbar \Omega -\beta\hbar \omega_c$ \cite{fock1928,darwin1931}, $\omega_c=\frac{eB}{2m}$, and $\Omega^2=\omega_0^2+\omega_c^2$. We consider moderate $B$ fields: the degeneracies $E_{n,l}$, for the same $l$, are lifted; but $E_{n,l}$ with different $n$ do not cross ($\omega_c/\omega_0\ll 1$). The single-particle eigenstates are grouped into ``atomic'' energy shells \cite{kouwenhoven2001}. The ground state $\psi_{0,0}$ is well separated from the first two excited states $\psi_{0,\pm1}$. 

We employ a description for few-electron DQDs that takes into account multiple energy levels and electron-electron interactions \cite{yang2011,wang2011}. As in the work of Burkard et al. \cite{burkard1999}, we construct a Hubbard model building upon the FD states. In contrast to more numerically oriented techniques, such an approach relies heavily on the chosen basis as couplings to other states are neglected. It has the advantage that all obtained results can be understood analytically. For two-electron DQDs, we include only the $\left(1,1\right)$, $\left(2,0\right)$, and $\left(0,2\right)$ electron configurations. The singlet (S) and $s_z=0$ triplet (T) states can be written as the product of spin and orbital parts:
$\Psi_{S/T}=
\left\{\phi_{1},\phi_{2}\right\}^{s/a}
\otimes\frac{\ket{\uparrow\downarrow}\mp\ket{\downarrow\uparrow}}{\sqrt{2}}
$. The electrons occupy states $\phi_{i}$, which need to be symmetrized/antisymmetrized for the S/T-state (as indicated by $\left\{\bullet,\bullet\right\}^{s/a}$).

In general, we cannot use a single FD state $\psi_{n,l}$ for the description of the states $\phi_{1/2}$ directly. But in the $\left(1,1\right)$ configuration, $\phi_{1/2}$ is close to the FD ground state $\psi_{0,0}^{L/R}$ on the left/right QD. In the $\left(2,0\right)$ and $\left(0,2\right)$ singlet configurations, both electrons fill the same orbital ground state, close to $\psi_{0,0}^{L/R}$ on the respective QD. For the triplet, the Pauli exclusion principle requires two different states to be occupied, so that one electron is in $\psi_{0,0}^{L/R}$ and the second electron is in $\psi_{0,1}^{L/R}$. As in atoms, the first electron shell $\psi_{0,0}^{L/R}$ is completed with two electrons in a singlet. We assume that in the six-electron configuration the first two electrons on each QD complete this first shell. We then adopt a frozen-core approximation: the $\left(3,3\right)$ configuration for six-electron DQDs is therefore equivalent to the $\left(1,1\right)$ configuration for two-electron DQDs [and similarly the $\left(4,2\right)$/$\left(2,4\right)$ and $\left(2,0\right)$/$\left(0,2\right)$ configurations]. One just needs to use the appropriate orbital wave function of these ``valence'' electrons. The valence orbital ground state is then $\psi_{0,1}^{L/R}$, while the first excited state is $\psi_{0,-1}^{L/R}$.

The two-electron DQD Hamiltonian is expressed in the basis $\left(1,1\right)_{S/T}$, $\left(2,0\right)_{S/T}$, and $\left(0,2\right)_{S/T}$ [and equivalently, without further specification, $\left(3,3\right)_{S/T}$,  $\left(4,2\right)_{S/T}$, and $\left(2,4\right)_{S/T}$ for the six-electron DQD]:
\begin{align}\label{C5-eq:2}
\mathcal{H}=\left(
\begin{array}{cccccc}
0 & 0 & \tau_S & 0 & \tau_S & 0\\
0 & 0 & 0 & \tau_T & 0 & \tau_T\\
\tau_S & 0 & U_S-\epsilon & 0 & 0 & 0\\
0 & \tau_T & 0 & U_T-\epsilon & 0 & 0\\
\tau_S & 0 & 0 & 0 & U_S+\epsilon & 0 \\
0 & \tau_T & 0 & 0 & 0 & U_T+\epsilon \\
\end{array}
\right).
\end{align}
The diagonal entries describe the energy of each state. The difference between the $\left(1,1\right)_{S}$ matrix element and $\left(1,1\right)_{T}$ matrix element is neglected, since it is usually small \cite{burkard1999}. Unequally occupied QDs are higher in energy by $U_{S/T}$. For simplicity we assume identical QDs on the left and the right. $\Delta\equiv U_T-U_S$ is the energy difference between the doubly occupied states. Electrostatic bias, modeled by the parameter $\epsilon$, influences the relative state energies of uniform and unequal electron arrangements. The off-diagonal elements in \eref{C5-eq:2} describe the spin-conserving hopping process of electrons between the dots.

\begin{figure}[htb]
\centering
\includegraphics[width=0.49\textwidth]{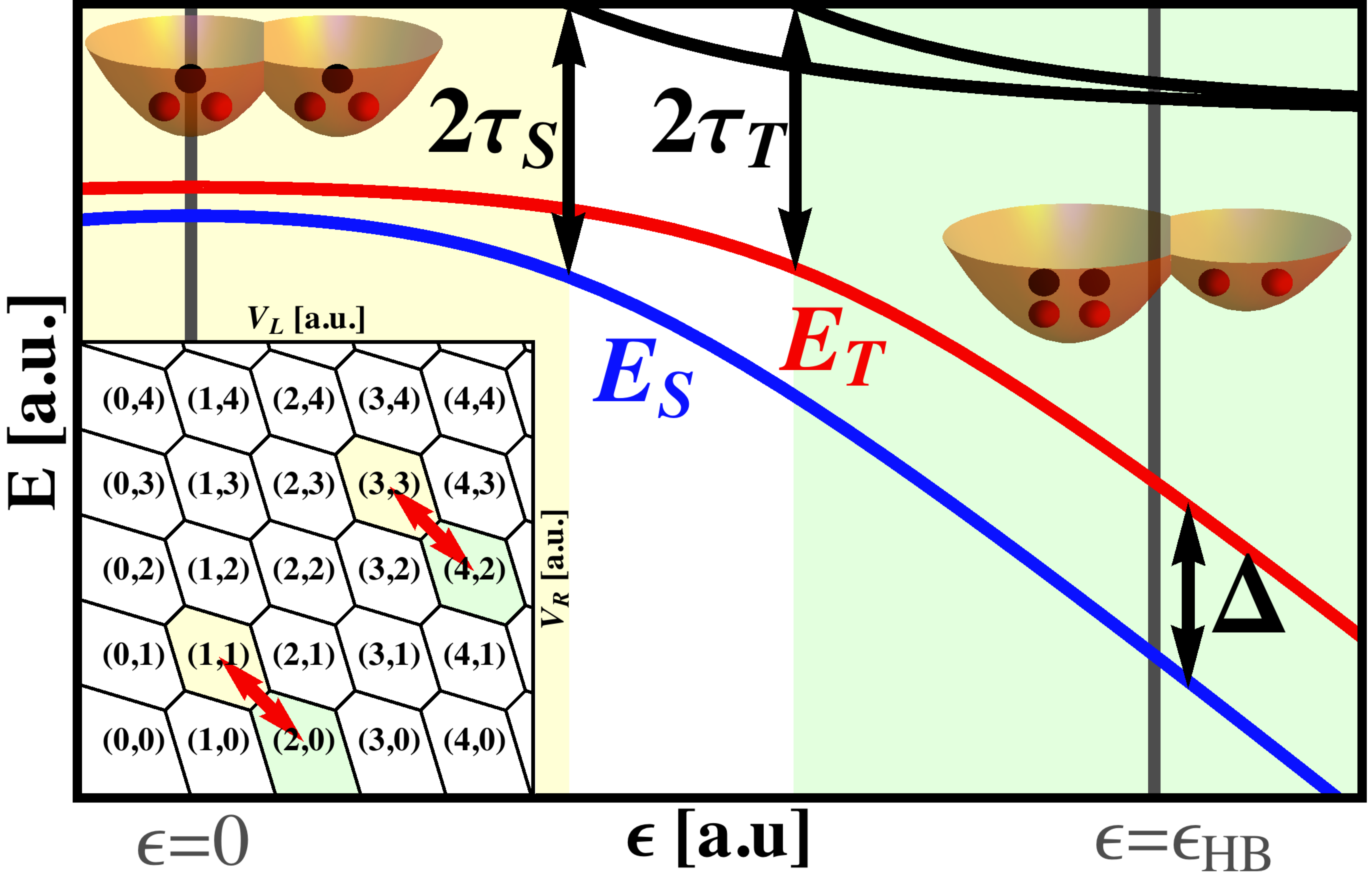}
\caption{Energy diagram for two- and six-electron DQDs, as described by \eref{C5-eq:2}. Electrostatic bias, modeled by $\epsilon$, transfers one electron from the uniform electron distribution on the two QDs towards two excess electrons on the left QD. The blue/red line represents the singlet/triplet ground state $E_{S/T}$; black curves are excited states. Charge noise generates fluctuations between $E_S$ and $E_T$, as described in the main text. The electron configurations are highly insensitive to charge noise at $\epsilon=0$ (the ``neutral'' electron configuration) and $\epsilon=\epsilon_{HB}$ (the ``high-bias'' configuration far away from the two anticrossings). The inset shows the charge stability diagram following van der Wiel et al. \cite{wiel2002} $\left(n_L,n_R\right)$ is the stable charge configuration for the left QD and the right QD. $V_{L/R}$ describes electrostatic voltages applied to the left/right QD; red arrows indicate gate tunings corresponding to the energy diagram.
\label{C5-fig:1}}
\end{figure}

\fref{C5-fig:1} shows the energy spectrum as a function of $\epsilon$. Close to state degeneracies $\left|\epsilon\right|=U_\sigma$, the hopping process hybridizes electron configurations of the same total spin. The ground state $E_S$/$E_T$ is shown in blue/red. At $\epsilon=0$, both energy levels are mainly in the $\left(1,1\right)$ charge configuration and their energy difference is minimal. $E_S$ and $E_T$ are lowered in energy for increasing bias due to the transfer of electrons between the QDs. For large $\epsilon$, the ground states are close to $\left(2,0\right)_{S,T}$ with an energy difference $\Delta$; we indicate one point deep in the $\left(2,0\right)$ region as the ``high-bias'' configuration $\epsilon=\epsilon_{HB}$.

Our treatment of the few-electron DQD is not self-consistent; it employs energy spectra of single-particle states, which are successively filled with electrons. The FD-states are a valid ansatz for the description of few-electron QDs if the electron-electron interaction influences the single-particle energies weakly or shifts all energy levels by a fixed value. The last scenario is consistent with the calculations of G\"u\c{c}l\"u et al., where the addition energy of interacting electrons has a constant offset compared to the noninteracting case (cf. \rcite{guclu2008}\footnote{A reasonable value of the constant $r_s$ in \rcite{guclu2008} for QDs that are used for quantum computation is $r_s=2$.}).
This is consistent with the constant-interaction model, introduced by Averin and Likharev \cite{averin1986,averin1991}, in which the energy spectrum of QDs remains unchanged when an electron is added to or removed from a QD.

%------------------------------------------------------------------------------------
%------------------------------------------------------------------------------------
%------------------------------------------------------------------------------------
\section{Charge Noise}
Charge noise is generally modeled by a random distribution of classical two-level fluctuators (TLFs), which couple electrostatically to QDs \cite{galperin2006,schriefl2006}. If the occupations of the charge traps (CTs) vary with a broad distribution of fluctuation rates, $1/f$ noise is generated. The coherence of the qubit decreases as seen by the time evolution of superpositions:
\begin{align}
\left\langle\sigma_x\right\rangle\left(t\right)=\left\langle e^{-i\int_0^t dt^\prime 
\frac{E_{ST}\left(t^\prime\right)}{\hbar}}\right\rangle\approx e^{-\left(\frac{t}{T_2}\right)^2}\sigma_x^{ideal}\left(t\right).
\label{C5-eq:3}
\end{align}
$E_{ST}\left(t\right)$ is the time-varying energy difference of the qubit levels, which deviates from the ideal value due to the coupling to TLFs:
$\delta E_{ST}= E_{ST}-\left\langle E_{ST}\right\rangle$.
$\left\langle\dots\right\rangle$ describes averaging over many experiments.
Assuming a static environment during one run, the coherence time $T_2$ is related to the statistics of the TLFs: $T_2^{-1}=\sigma_{\delta E}/\sqrt{4\pi}$; $\sigma_{\delta E}^2=\int_{-\infty}^{\infty}dt~ \delta E^2\left(t\right)$ \cite{nakamura2002,makhlin2004,ithier2005,hu2006}.

We analyze the relative energy shift of the qubit levels of a DQD which couples to a CT. In first-order perturbation theory, the fluctuation of the singlet-triplet splitting is described by \cite{ramon2010}:
\begin{align}
\delta E_{ST}^{\left(1\right)}=
\Dirac{\Psi_T}{e\Phi_{CT}}{\Psi_T} - 
\Dirac{\Psi_S}{e\Phi_{CT}}{\Psi_S}.
\label{C5-eq:4}
\end{align}
$\Phi_{CT}$ is the electrostatic potential of a CT. Since for QDs that are suitable for qubits CTs are at some distance from the QD center, we make a multipole expansion of $\Phi_{CT}$:
$\Phi_{CT}\approx \Phi\left(\mathbf{r}_0\right)-\mathbf{E}\left(\mathbf{r}_0\right)\cdot\mathbf{r}-\frac{1}{2}\left[\partial_i E_j\left(\mathbf{r}_0\right)\right]r^i r^j$ \cite{jackson1999}. $\mathbf{r}_0$ is the position of the CT relative to the center of the DQD, $\mathbf{r}$ is the QD electron coordinate. This expansion resolves the coupling of a TLF into 
dipole ($-\mathbf{E}\cdot \mathbf{d}$) and 
quadrupole [$-(1/2)\left(\partial_i E_j\right) \cdot  \mathcal{Q}^{ij}$] terms; 
$d^i=e\Dirac{\Psi}{r^i}{\Psi}$, and $\mathcal{Q}^{ij}=e\Dirac{\Psi}{r^ir^j}{\Psi}$ are the first two electric moments of the DQD. We analyze two points in the charge stability diagram $\epsilon=0$ and $\epsilon=\epsilon_{HB}$ (``sweet spots'', introduced in \fref{C5-fig:1}) at which coupling is weak to TLFs. High couplings are obtained if the qubit states have different dipole moments, which generate energy shifts scaling like, e.g., $1/r_0^2$.

The eigenstates of the singlet-triplet qubit of \eref{C5-eq:2} can be approximated at $\epsilon=0$:
$
\Psi_{S/T} \propto \ket{11}_{S/T} -\frac{\tau_{S/T}}{U_{S/T}} \left(\ket{0,2}_{S/T}+\ket{2,0}_{S/T}\right)
$.\footnote{We assume two identical QDs for simplicity. However, one can also finds a sweet spot for distinct QDs. Then $\Psi_{S/T}$ has different weights in $\left(2,0\right)$ and $\left(0,2\right)$ at the $\left(1,1\right)$-sweet spot; additionally, the sweet spot is not in the center of a charge stability region. Charge traps couple dominantly to DQDs via their electric field, which induces a small shift in the bias parameter $\epsilon$. This mechanism generates no dephasing at the minimum of $\left[E_{T}-E_{S}\right]\left(\epsilon\right)$. We note that it might well be desirable to use one weakly and one strongly confined QD in future experiments. A strongly confined QD simplifies qubit initializations, while the large QD can be used for exchange manipulations.}
These states have equivalent dipole moments for the two qubit levels; the charge distribution of a DQD arranged in x-direction has mirror symmetry to the y-z plane. The quadrupole contribution describes the spread of the charge distributions. The unequal degree of hybridization of the singlet state and the triplet state creates different variances in x-direction: 
$\delta E_{ST}^{\left(1\right)}$
$\approx$
$\left[\left(\frac{\tau_S}{U_S}\right)^2-\left(\frac{\tau_T}{U_T}\right)^2\right]$
$\cdot$
$\frac{e d_0^2}{4}$
$\cdot$
$\frac{e}{4\pi\epsilon_0\epsilon_r}$
$\left(\frac{x_0^2}{r_0^5}-\frac{1}{r_0^3}\right)$. The first factor describes the hybridizations for the singlet and the triplet, the second factor involves the interdot distance $d_0$ of the DQD, and the third factor is the gradient contribution of the electric field of the TLF. It describes an energy shift proportional to the hybridization of the ground state $\left(\tau_{S/T}/U_{S/T}\right)^2$, which decays like $1/r_0^{3}$ in the TLF-QD distance. A similar expression holds for the six-electron DQD.

Considering the two-electron DQD for high bias ($\epsilon=\epsilon_{HB}$), the left QD is lower in energy than the right QD. We assume we are far away from the transition region in which the valence electrons occupy single-particle eigenstates of the left QD. For the singlet both electrons are placed into $\psi^{L}_{0,0}$; for the triplet one electron occupies $\psi^{L}_{0,0}$, the other $\psi^{L}_{0,1}$. The dipole contributions to \eref{C5-eq:4} vanish because the charge distribution of $\Psi_{S}$ and $\Psi_{T}$ are both centered at the left QD. The quadrupole contribution of \eref{C5-eq:4} is: $\delta E_{ST}^{\left(1\right)}\approx
\left(\frac{e\hbar}{4m\Omega}\right)
\cdot\left(\frac{e}{4\pi\epsilon_0\epsilon_r}\right)
\left(\frac{x_0^2+y_0^2}{r_0^5}-2\frac{z_0^2}{r_0^5}\right)$. The first factor comes from the different spread of the density of the qubit states, while the second factor describes the influence of the CT. We find a $1/r_0^3$-scaling in the CT-QD distance as for the low-bias sweet spot.

The situation improves for six-electron DQDs. As the valence electrons' wave functions $\psi_{0,\pm1}$ are complex conjugates of each other, not only the quadrupole term of \eref{C5-eq:4}, but {\it all multipole contributions vanish}. $\delta E_{ST}^{\left(1\right)}$ depends only on the charge density of the single-electron wave functions, as $e\Phi_{CT}$ in \eref{C5-eq:4} contains exclusively single-particle operators. The second-order dipole contribution of TLFs (second-order Stark effect) vanishes accordingly, since it involves only an overall shift of the confining potential. The first nonvanishing contributions are second-order quadruple couplings:
$\delta E_{ST}^{\left(2\right)}\approx
-\frac{e}{2}\left(\partial_i E_j\right)^2\frac{\left|\Dirac{\psi_{0,1}}{r^i r^j}{\psi_{0,-1}}\right|^2}{E_{0,1}-E_{0,-1}}
$. We note that this contribution has $1/r_0^6$ scaling with the CT-QD distance, which suppresses $\delta E_{ST}^{\left(2\right)}$ considerably. This protection criterion for six-electron DQD is strongest for perfect circular symmetry. For weakly elliptic QDs, $V=\frac{m\omega_0^2}{2}\rho^2\left[1+\beta\cos\left(2\phi\right)\right]$, the diagonal terms of the quadrupole tensor differ, weighted by the ellipticity $\beta$:
$\mathcal{Q}_{\phi_{0,1}}^{xx,yy}-\mathcal{Q}_{\phi_{0,-1}}^{xx,yy}
\approx\mp\beta\frac{3\hbar}{m\omega_c}+\mathcal{O}\left(\left(\frac{\omega_c}{\omega_0}\right)^2\right)
$, giving a small $1/r_0^3$ contribution.

A summary of $\delta E_{ST}$ is given in \tref{C5-tab:1}. For $\epsilon=0$, a sweet spot is present for both the $\left(1,1\right)$ and $\left(3,3\right)$ cases. $\delta E_{ST}$ comes from a direct coupling of TLFs to the quadrupole moment of the DQD. The energy shifts are on the order of a few gigahertz, corresponding to a dephasing time of $\text{ns}$. This time scale is consistent with experiments on DQD charge qubits \cite{petta2004,petersson2010}. Another sweet spot is identified at $\epsilon=\epsilon_{HB}$. For two-electron DQDs the scaling in $r_0$ is identical to $\epsilon=0$, only lacking the hybridization factor $\left(\tau_{S/T}/U_{S/T}\right)^2$. $T_2$ is improved for six-electron DQDs, as the CTs modify $E_{ST}$ coupling only to the quadrupole moment in second order.

\begin{table}[htb]
\centering
\begin{tabular}{c  c  c c c }
\hline
\hline
&$\epsilon=0$&\multicolumn{3}{c}{$\epsilon=\epsilon_{HB}$}\\
& two and six electrons & two && six\\
  \hline
  Mechanism & \multicolumn{4}{c}{Coupling to electric quadrupole moment}\\
  & direct & direct &/& second order\\
  Scaling &	$\sim \left(\frac{\tau_{S/T}}{U_{S/T}}\right)^2\frac{1}{r_0^3}$	&$\sim \frac{1}{r_0^3}$  && $\sim \frac{1}{r_0^6}$\\
  $T_2$ &	$\sim \text{ns}$	&$< \text{ns}$	&&$>\text{ns}$\\
\hline
\hline
\end{tabular}
\caption{Influence of CTs on two- and six-electron DQDs. $E_{ST}$ is shifted, depending on the distance $r_0$ between CT and DQD. Two sweet spots $\epsilon=0$ and $\epsilon=\epsilon_{HB}$ are identified (cf. \fref{C5-fig:1}). The hybridization factor $\left(\tau_{S/T}/U_{S/T}\right)^2$ [parameter introduced in \eref{C5-eq:2}] enhances the coherence time for $\epsilon=0$. $\delta E_{ST}$ decreases with $r_0$.  Note that for the six-electron DQD the decay is much faster at $\epsilon=\epsilon_{HB}$: the CTs and the qubit couple only in second-order perturbation theory.
\label{C5-tab:1}}
\end{table}

%------------------------------------------------------------------------------------
%------------------------------------------------------------------------------------
%------------------------------------------------------------------------------------
\section{Robust Single-Qubit Gating}
\begin{figure}[htb]
\centering
\includegraphics[width=0.49\textwidth]{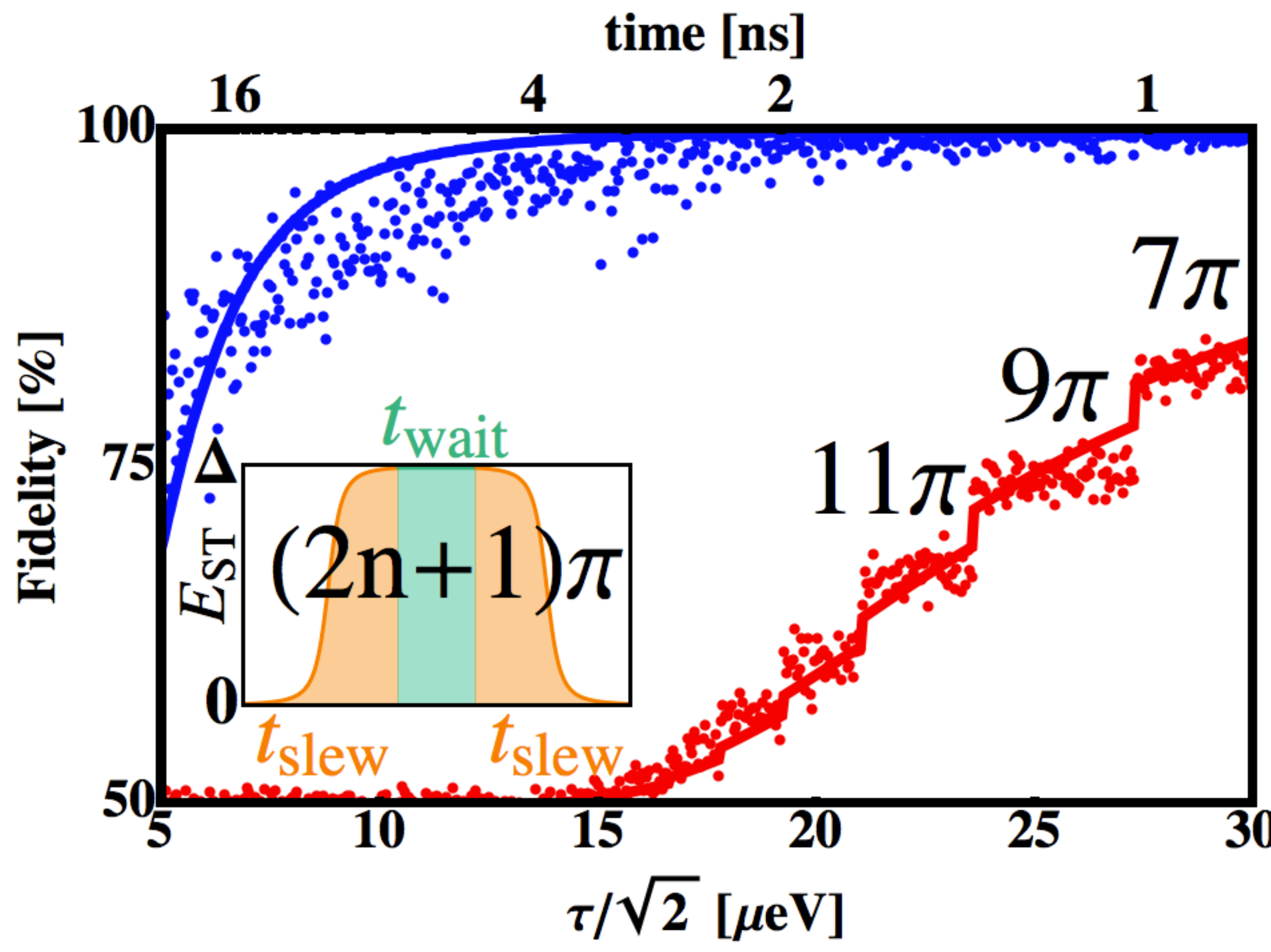}
\caption{Fidelity analysis for $\pi$-phase gate for a two/six-electron DQD, shown in red/blue. Points are from simulations involving $1/f$ noise sources. The fidelities are poor for slow manipulation times, which are required by small tunnel couplings $\tau=\tau_S=\tau_T$ (cf. the transition probability at a Landau-Zener crossing $P_{LZ}$). Increasing $\tau$ allows faster qubit manipulations, which increases the fidelity. The fidelity of the six-electron DQD approaches $1$, while it stays much lower for the two-electron system. The solid lines are fits using \eref{C5-eq:3}, with $T_2=1.5/29.3$ ns for the two/six-electron system.
The inset describes the pulse profile of a $\pi$-phase gate. Starting from $\epsilon=0$, the DQD is biased to $\epsilon=\epsilon_{HB}$; we linearly increase $\epsilon$ for a time $t_{\text{slew}}$. The qubit stays at $\epsilon=\epsilon_{HB}$ for $t_{\text{wait}}$; finally the qubit is brought back to $\epsilon=0$, picking up in total an odd number of $\pi$ rotations. The overall gate time equals $2t_{\text{slew}}+t_{\text{wait}}$.
\label{C5-fig:2}}
\end{figure}

We have identified two points $\epsilon=0$ and $\epsilon=\epsilon_{HB}$ that are well isolated from external noise sources. It is possible to manipulate the qubit while staying mainly at these sweet spots. Changing the magnitude of $E_{ST}$ produces a phase gate: $\mathcal{U}=J\sigma_z$, $J=\int_{0}^{t}d\tau~E_{ST}\left(\tau\right)$.  $E_{ST}$ is small at $\epsilon=0$, while $E_{ST}=\Delta$ at $\epsilon=\epsilon_{HB}$. A possible gate sweep starts from $\epsilon=0$ and tunes the bias rapidly to $\epsilon=\epsilon_{HB}$; after some waiting time the bias is brought back to $\epsilon=0$ (cf. inset of \fref{C5-fig:2}). While the manipulation must be fast to avoid charge noise, it should still be adiabatic with respect to the coupling to excited states (cf. \fref{C5-fig:1}). The slew rate is limited by the leakage to higher states, which is approximated with the transition probability at a Landau-Zener crossing of strength $\tau$ which is crossed with velocity $v_{\text{slew}}$ \cite{shevchenko2010}: 
$P_{LZ}=e^{-2\pi\frac{\tau^2}{\hbar v_{\text{slew}}}}$.
Since the tunnel coupling enters $P_{LZ}$ quadratically, realistic values of $\tau$ allow very fast manipulations with permitted pulse lengths far below nanoseconds.

We show a fidelity analysis of a $\pi$-phase gate for a two- and six-electron DQD in \fref{C5-fig:2}. The slew rates are fixed through $P_{LZ}$ to produce negligible leakage (cf. \aref{C5-app:DescFidAn} for further information about the specific parameter choice and the setup of the simulation). We use similar densities of the CTs for the two- and six-electron DQDs, which are positioned randomly around the DQD to generate $1/f$ noise; the coupling to CTs vary the parameter $\Delta$ through electrostatic couplings to the DQD potential. We exclude a volume around the QD, where no CTs are permitted; such nearby TLFs make the DQD completely nonfunctional as a qubit. We take the excluded volume for two-electron DQDs to be considerably larger than for the six-electron system. Fluctuations in the 
 coupling $\tau$ or the pulse profiles are disregarded. The sweet spots, especially $\epsilon=\epsilon_{HB}$, offer the advantage that $E_{ST}$ does not change over a wide range of $\epsilon$.

The fidelity of the gate, both for the two-electron and the six-electron systems (blue/red), is low for small tunnel couplings $\tau$. The fidelity increases very quickly with $\tau$ for six-electron DQDs and reaches an ideal value very close to $1$. The improvement of the fidelity for the two-electron system is much slower. We approximate the curves according to \eref{C5-eq:3}, yielding a coherence time of $1.5$ ns for the two-electron system and $29.3$ ns for the six-electron case. Steps seen for the two-electron system are generated by different waiting times in $\left(2,0\right)$ when constructing a $\pi$-phase gate; a one-parameter fit to \eref{C5-eq:3} cannot completely reproduce these results.

%------------------------------------------------------------------------------------
%------------------------------------------------------------------------------------
%------------------------------------------------------------------------------------
\section{Conclusion}
We propose a fast and robust way to manipulate singlet-triplet qubits (STQs) via a high-bias phase gate. Contrary to current realizations of phase gates, our approach works by going to high bias. The qubit couples to CTs weakly, and we manipulate rapidly between two sweet spots. The ``high-bias'' sweet spot $\epsilon_{HB}$ is not at a specific point in the charge diagram; there is a large range of parameters where $E_{ST}$ is constant. 
Note that the Rabi rotation gate needed for full qubit control is envisioned to occur also at a sweet spot (at $\epsilon=0$), employing magnetic field gradients.
It is worth pointing out that the proposed high-bias phase gate works also as a maximally entangling two-qubit gate for single-QD qubits \cite{loss1998}.  

It would be favorable for our proposal that DQDs have small singlet-triplet energy splitting at $\epsilon=\epsilon_{HB}$ ($\Delta$, cf. \fref{C5-fig:1}), to give comfortable electrical manipulation times (subnanosecond has become accessible \cite{cao2013}). DQDs with $\Delta$ on the order of $30$ GHz have been reported \cite{dial2013}. One can decrease the singlet-triplet energy splitting further by using favorable dot sizes and external magnetic field parameters. Indeed, we note that a transition from a singlet to a triplet ground state is indicated in calculations on four-electron QDs \cite{kyriakidis2005}. However, a triplet ground state is not permitted in our parameter regime of moderate magnetic fields and for QDs with weak state renormalizations from Coulomb interactions.

A clear prediction of our work is that the many-electron QDs, specifically those for which the valence electrons occupy the second shell, are uniquely suited to protect STQs from charge noise because singlet and triplet charge densities are identical in the second shell. The manipulation of our six-electron STQs can be performed in the same way as for the two-electron DQDs, including initialization, manipulation, and measurement.  Additional noise sources, which couple in via the charge density, like pure phonon dephasing \cite{hu2011,gamble2012}, are also directly suppressed in our approach. 
We are hopeful that the prospect of an order of magnitude improvement in gate fidelity will motivate the further experimental exploration of the multielectron regime in QD qubits.

\textit{Acknowledgments} \textthreequartersemdash\ 
We acknowledge fruitful discussions with H. Bluhm, G. Burkard, and C. M. Marcus. We are grateful for support from the Alexander von Humboldt foundation.

%------------------------------------------------------------------------------------
%------------------------------------------------------------------------------------
%------------------------------------------------------------------------------------
\begin{appendix}
%------------------------------------------------------------------------------------
%------------------------------------------------------------------------------------
%------------------------------------------------------------------------------------
\section{Description of the Fidelity Analysis
\label{C5-app:DescFidAn}}
\begin{figure}[htb]
\centering
\includegraphics[width=0.49\textwidth]{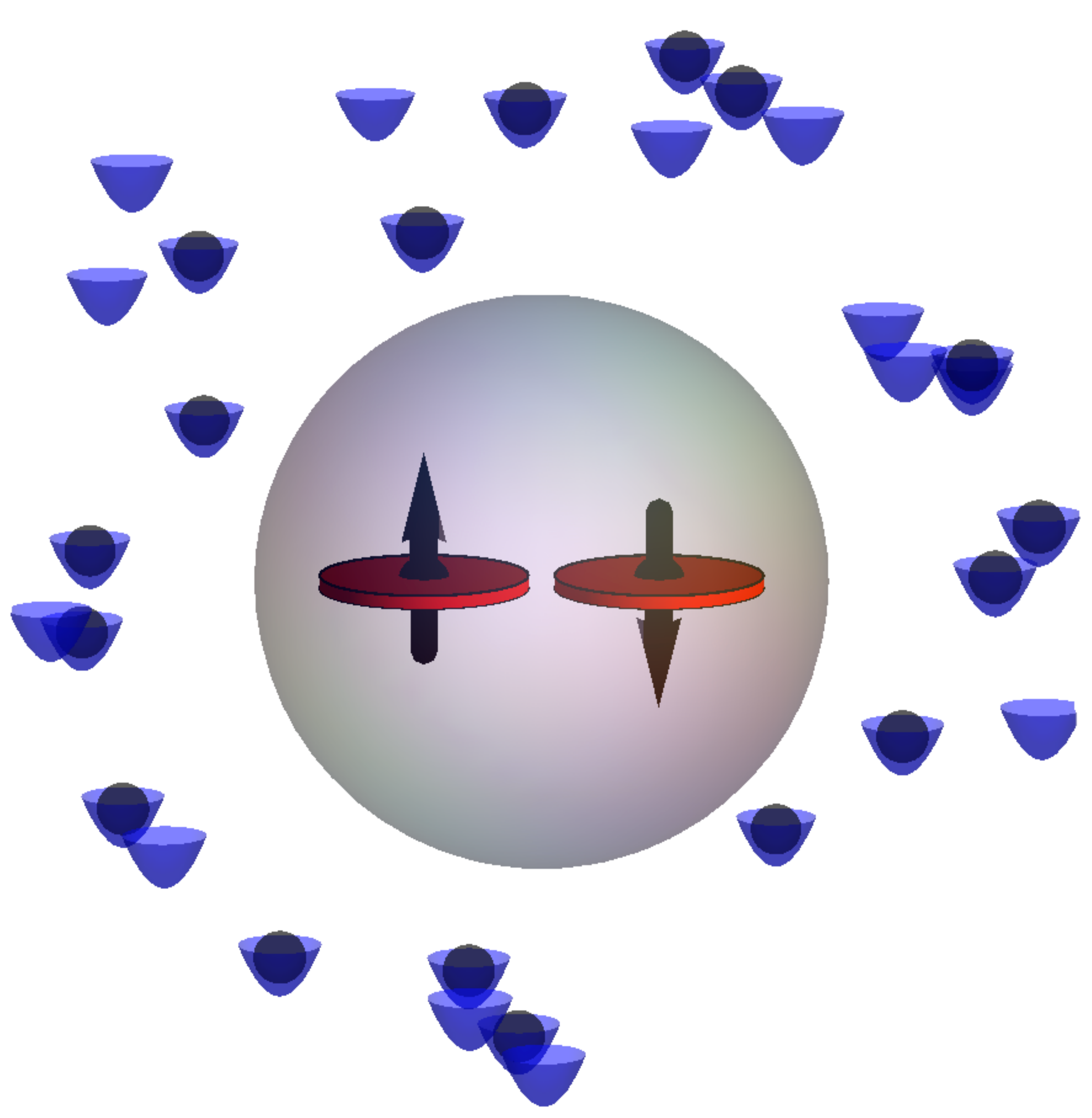}
\caption{Model of charge traps which couple to a DQD electrostatically. No charge traps are permitted in a volume surrounding the QD; charge traps in this area make the qubit completely nonfunctional. All charge traps fluctuate between being filled with one electron or being empty; they are randomly generated for a single run of the simulation. As charge noise has dominant spectral weight at low frequencies (i.e. lower than the inverse gate time), the electrostatic potential of the charge traps is kept constant during one gate simulation.
\label{C5-app:fig:1}}
\end{figure}

We model charge noise acting on DQDs by a random distribution of charge traps, being either filled or empty (cf. \fref{C5-app:fig:1}). The time evolution during an exchange gate is determined numerically using quantum process tomography \cite{nielsen2000}. We generate a distribution of TLFs with a broad range of switching rates $\gamma$ for each run of the simulation. A reasonable probability distribution is $P\left(\gamma\right)\sim1/\gamma$ \cite{burkard2009}. The charge distribution is constant during one run of the simulation, while the potential fluctuates between successive simulations. This scenario mimics consecutive measurements with a long time between the measurements. 

The coupling strength of the DQD and a TLF is determined by their distance. As described in the main text, we take the shift in the singlet-triplet splitting $\delta E_{ST}$ as the only dynamic variable. For the $\left(2,0\right)$ and $\left(4,2\right)$ configurations, the energy shifts are:
\begin{widetext}
\begin{align}
\label{C5-app:eq:1}
\delta E_{ST}^{\left(2,0\right)}=&
\left(\frac{e\hbar}{4m\omega_0}\right)
\left(\frac{e}{4\pi\epsilon_0\epsilon_r}\right)
\left(\frac{x_0^2+y_0^2}{r_0^5}-2\frac{z_0^2}{r_0^5}\right)
+\mathcal{O}\left[\left(\omega_c/\omega_0\right)^2\right],\\
\delta E_{ST}^{\left(4,2\right)}=&\frac{9}{16}\left(\frac{\hbar}{m^2\omega_0^3}\right)\left(\frac{\omega_c}{\omega_0}\right) 
\left(\frac{e^2}{4\pi\epsilon_0\epsilon_r}\right)^2\left(\frac{\left(x_0^2+y_0^2\right)^2}{r_0^{10}}\right)
+\mathcal{O}\left[\left(\omega_c/\omega_0\right)^2\right].
\end{align}
\end{widetext}
The excessively occupied QD is positioned in the x-y plane at the coordinate origin, while charge traps occupy the space around the DQD.

We use material parameters of GaAs. The confining strength $\hbar \omega_0=3~\text{meV}$ is a common approximation for QDs \cite{burkard1999}. $\omega_c/\omega_0=0.1$ describes moderate external magnetic fields of $0.7~T$. The singlet-triplet splitting $\Delta$ is rather small, consistent with Dial et al. \cite{dial2013}. All parameters are summarized in \tref{C5-tab:1}.

The electron distribution can be approximated by the spread of the ground state wave function: $a_B\equiv\sqrt{\frac{\hbar}{m\omega_0}}\approx 20~\text{nm}$. We use $250$ TLFs with a distance $\left[2.5,15\right]a_B$ from the coordinate origin for the six-electron system. For the two-electron system, we need to exclude a larger volume around the DQD. Otherwise the energy shifts due to \eref{C5-app:eq:1} destroy the qubit fidelity completely. To generate the same density of TLFs around the DQD, we include $196$ charge traps with a distance $\left[15,25\right]a_B$ from the origin.

\begin{table}[htb]
\centering
\begin{tabular}{l c c}
\hline
\hline
\multicolumn{2}{c}{Parameter} & Value\\
\hline
energy difference between & \multirow{2}{*}{$U_S$} & \multirow{2}{*}{$0.5~\text{meV}$}\\
$\left(1,1\right)/\left(2,0\right)$ or $\left(4,2\right)/\left(3,3\right)$	&&\\
singlet-triplet splitting	&	$\Delta$	&	$10~\text{GHz}$\\ &&\\
dielectric constant	&	$\epsilon_r$ & $12.5$\\
effective mass	&	$m$	&	$0.067 m_e$\\
confining energy	&	$\hbar \omega_0$	&	$3~\text{meV}$\\
magnetic energy	&	$\hbar \omega_c$	&	$0.1\hbar\omega_0$\\
\hline
\hline
\end{tabular}
\caption{Parameters used for the simulations of STQs. $U_S$ and $\Delta$ are chosen to describe the DQD dynamics according to \eref{C5-eq:2} in the main text. The dielectric constant $\epsilon_r$ and effective mass $m$ correspond to GaAs; $\omega_0$ and $\omega_c$ mimic common confining strengths and magnetic fields of $0.7~T$.
\label{C5-app:tab:1}}
\end{table}
\end{appendix}
%------------------------------------------------------------------------------------
%------------------------------------------------------------------------------------
%------------------------------------------------------------------------------------

%------------------------------------------------------------------------------------
%------------------------------------------------------------------------------------
%------------------------------------------------------------------------------------
\bibliography{library}
\end{document}